\documentstyle[twoside,epsfig]{article}

\input{ijmpal.tex}

\def\sinphi{$\sin 2\phi_1\ $}

\begin{document}

\runninghead{Measurement of \sinphi at Belle} {Measurement of \sinphi at Belle$\ldots$}

\normalsize\textlineskip
\thispagestyle{empty}
\setcounter{page}{1}


\vspace*{0.88truein}

\fpage{1}
\centerline{\bf MEASUREMENT OF sin(2$\phi_1$) AT BELLE}
\vspace*{0.035truein}
\centerline{\footnotesize JORGE L. RODRIGUEZ}
\centerline{\footnotesize Representing the Belle Collaboration}
\vspace*{0.015truein}
\centerline{\footnotesize\it Department of Physics and Astronomy, University
of Hawaii, 2505 Correa Road}
\baselineskip=10pt
\centerline{\footnotesize\it Honolulu, Hawaii 96822,
USA}

\vspace*{0.21truein}
\abstracts{With 6.2 $fb^{-1}$ of data collected on the $\Upsilon (4S)$,
Belle reports its first measurement of 
$\sin 2\phi_1 = 0.45^{+0.43}_{-0.44}({\rm stat})^{+0.07}_{-0.09}({\rm sys})$.
The result was obtained by fitting the proper time
 distribution of flavor tagged and fully reconstructed neutral $B$
mesons decays to five different charmonia plus a $K_s$ or $K_L$ channels. In 
this paper the analysis and results will be described briefly.}{}{}

\vspace*{1pt}\textlineskip	
\vspace*{-0.5pt}

\textheight=7.8truein
\setcounter{footnote}{0}
\renewcommand{\thefootnote}{\alph{footnote}}

\section{Introduction}
\noindent
In the Standard Model, the violation of the $CP$ symmetry by the weak
interaction is possible via the complex phase in the CKM mass mixing
matrix. This phase can be extracted by observing processes that
involve flavor transitions spanning the three quark generations. In
this analysis we consider decays of the neutral $B$ meson to final
states which are $CP$ eigenstates. The interference between the weak
phase in direct decays and decays that proceed via mixing leads to an
asymmetry in the time-dependent decay rate which is exploited here to
measure the value of \sinphi,($\sin 2\beta$). The particular decay
modes considered were selected because they have relatively large
branching fractions, clear experimental signatures and are essentially
free from theoretical uncertainties. Five neutral $B$ decay channels
were used: $B\rightarrow J/\psi K_S, \psi(2S)K_s$, $\chi_{c1}K_s$,
$J/\psi K_L$ and $J/\psi\pi^0$.  The first three channels have odd
$CP$ while the last two have even $CP$.

\section{Analysis Procedure}
\noindent
Our analysis includes all of the available data collected by the Belle
experiment$^1$ up through the end of the Summer 2000. The data sample
consists of 6.2 $fb^{-1}$ taken at the $\Upsilon(4S)$. At Belle, the
center of mass of $B\bar{B}$ pair system is boosted by the asymmetric
configuration of the $e^+e^-$ beams. The boost ($\gamma\beta =
0.425$), the long-lifetime of the $B$ meson and the excellent
resolution of the silicon-vertex detector makes possible measurements
of time-dependent decay rates from the displacement of the
reconstructed $B$ vertices in the boost direction. To deduce the
flavor of the $B_{CP}$ meson, its decays to a $CP$-eigenstate is
flavor non-specific, we tag the flavor of the $B_{tag}$, the other $B$
meson\footnote{Since, the decay of the $\Upsilon(4s)$ creates a pair
of $B\bar{B}$ mesons in a coherent quantum state, tagging the flavor
of one of the $B$ automatically determines the flavor of the other $B$
at the time when the tagged $B$ decayed.}. The relationship between
the time-dependent decay rate, in terms of the proper-time $\Delta t =
t_{CP}-t_{tag}$, and the $CP$ violating parameter is given by,
\begin{equation}
{dN\over d \Delta t} \left(B\rightarrow f_{CP}\right) \propto 
e^{-\Gamma |\Delta t| }
\{ 1 - (1-2\omega)\eta_{CP}\sin 2\phi_1\sin(\Delta m\Delta t)\},
\end{equation}
here $\Delta m$ determines the frequency of $B^0\bar{B}^0$ mixing, 
$\eta_{CP}$ is either $+1(-1)$ for even(odd) $CP$ final states and 
\sinphi is the $CP$-violating term. The $(1-2\omega)$ term,
known as the dilution factor, accounts for the possibility of 
mis-assigning the flavor to the $B_{tag}$ (wrong-tag). 

The analysis procedure used to extract the value of \sinphi includes
four main components: (1) reconstruct, exclusively, the decay of the
$B_{CP}$ meson and identify candidate events, (2) determine the flavor
of $B_{CP}$, at the decay time of $B_{tag}$ by tagging its flavor, (3)
determine the proper-time of the decay by measuring the difference
between the $z$ decay vertices of the two $B$ mesons and compute the
proper-time from the $\Delta t = \Delta z/\gamma\beta$ relation and
(4) form the proper-time distribution from the sample of tagged and
fully reconstructed events and extract the value of $\sin 2\phi_1$
from an unbinned maximum likelihood fit.

\subsection{Reconstruction of the $B_{CP}$ decay $^{2,3}$}
\begin{figure}[t]
\vspace*{13pt}
\centerline{\epsfig{file=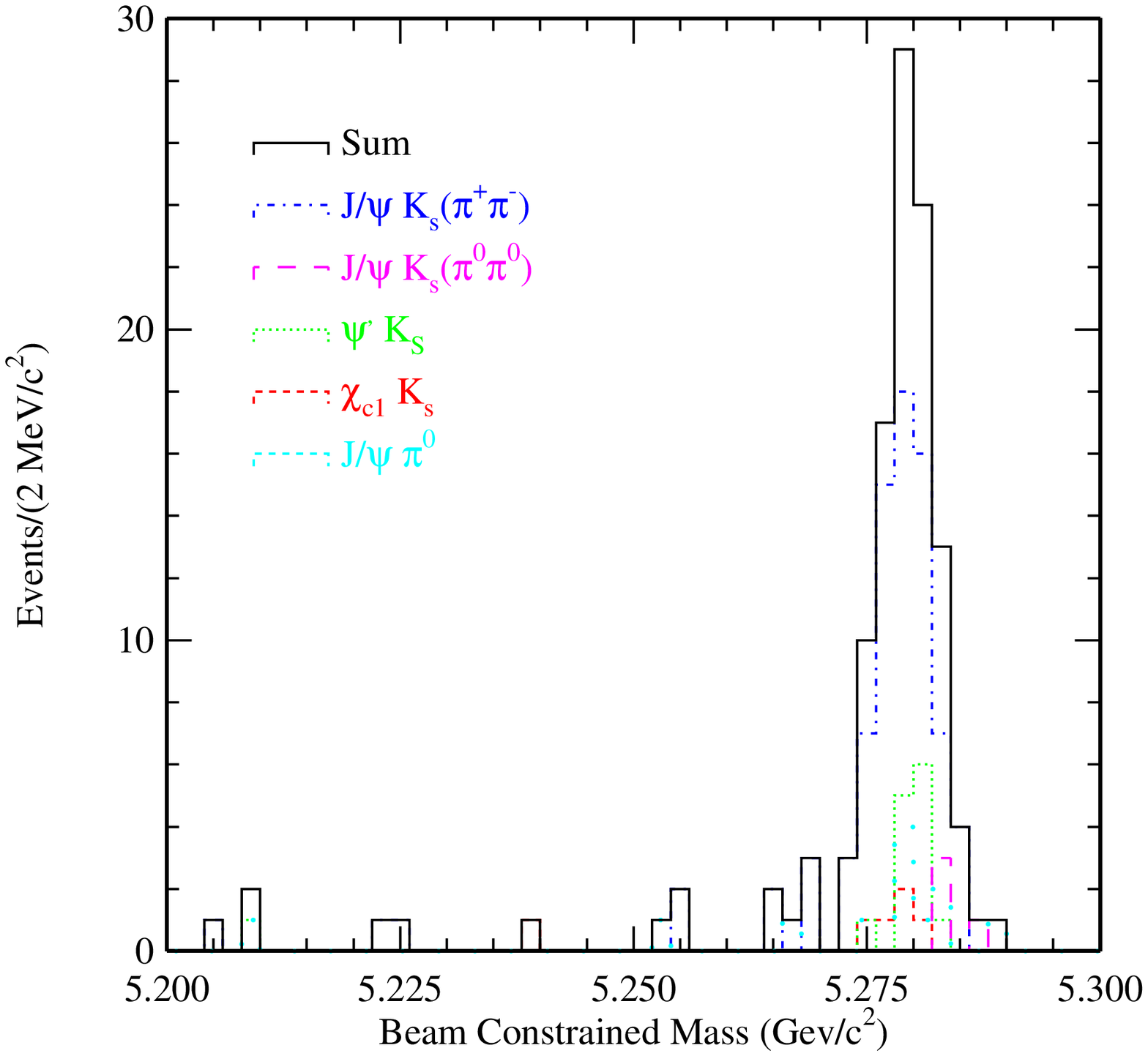,height=2.36in}\epsfig{file=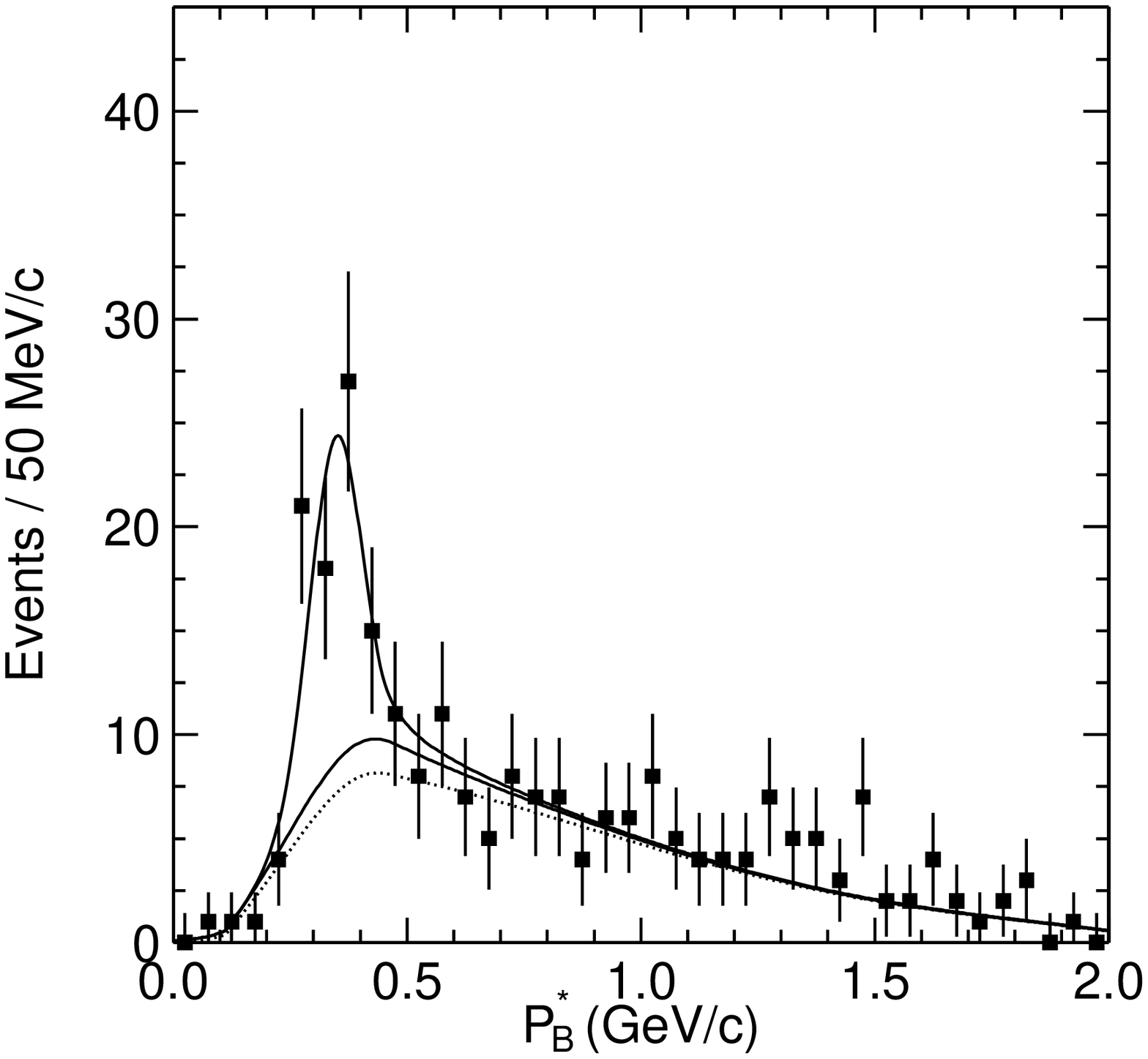,height=2.5in}}
\fcaption{The plot on the left, shows the beam-constrained mass
distribution used to identify the number of reconstructed $B_{CP}$ in all but the
$J/\psi K_L$ mode. The plot on the right, shows the $p_{B}^*$
distribution used to identify the $J/\psi K_L$. The signal region in $p_{B}^*$ is
defined from 0.2 to 0.45 GeV/c. The background distribution in the fit is divided into components describing the contribution from resonant and (non-resonant) $B\rightarrow J/\psi K^{*0}(K_L\pi^0)$ and all other $B$ decays.}
\label{jrod:fig:signals}
\end{figure}

\noindent
Event shape variables were used to reject continuum background and
particle identification requirements were imposed on tracks used to
reconstruct the decay products of the $B_{CP}$. To form $J/\psi$ and
$\psi(2S)$ candidates we used dilepton channels ($\mu^+\mu^-,e^+e^-$),
correcting for final-state radiation in the electron channel. For
$\psi(2S)$ candidates we also used the $J/\psi\pi^+\pi^-$ mode. For
$\chi_{c1}$ candidates we used only the $J/\psi\gamma$ decay channel.
$K_s$ candidates were selected from among $\pi^+\pi^-$ and
$\pi^0\pi^0$ combinations. The neutral mode was used exclusively in
the reconstruction of $B\rightarrow J/\psi K_s$. The signals for $CP$
eigenstates were identified kinematically via the beam-constrained
mass and $\Delta E$ (variables in which the beam-energy is substituted
for the measured energy) distributions, see
Fig. \ref{jrod:fig:signals}. The total number of $B$ candidates found
for each mode are listed in Table \ref{jrod:tbl:Bnum}.

\begin{table}[b!]
\tcaption{Number of $B_{CP}$ candidates reconstructed and tagged}
\label{jrod:tbl:Bnum}
\centerline{\begin{tabular}{|c|l|c|c|c|}
\hline
\hline
\bf \it CP & \bf{{\it B} Decay Mode} & \bf{Signal} & \bf{Background} & \bf{Tagged}\\
\hline
$-1$&$J/\psi K_s, K_s\rightarrow \pi^+\pi^-$             & 70 & 3.4 &40\\
$-1$&$J/\psi K_s, K_s\rightarrow \pi^0\pi^0$             & 4  & 0.3 & 4\\
$-1$&$\psi(2s)K_s, \psi(2s)\rightarrow l^+l^-$           & 5  & 0.2 & 2\\
$-1$&$\psi(2s)K_s, \psi(2s)\rightarrow J/\psi\pi^-\pi^+$ & 8  & 0.6 & 3\\
$-1$&$\chi_{c1}K_s$                			 & 5  & 0.8 & 3\\\hline
$+1$&$J/\psi K_L  $                			 &102 &48.0 &42\\
$+1$&$J/\psi\pi^0 $                			 & 10 & 0.6 & 4\\\hline
& Total				                         &204 &53.9 &98\\\hline 
\hline
\end{tabular}}
\end{table}
\subsection{Flavor Tagging $^3$}
\noindent
To determine the flavor of the $B_{CP}$ candidates, we examined the
remaining tracks in the event to identify the flavor of $B_{tag}$. The
flavor tagging algorithm employs four methods: (1) high momentum
($p_l^*>1.1{\rm GeV/c}$) lepton charge; this tags the $b$ quark flavor
via its primary decay to a lepton, (2) sum of the charge for all well
identified kaons; this tag relies on the flavor of the $s$ quark from
cascade decays, (3) medium momentum lepton ($0.6<p_l^*<1.1$ GeV/c);
here if the sum of $p_l^*$ and $p_{miss}^*$ is greater than 2.0 GeV/c
then the charge of the lepton tags the $b$ quark flavor as in (1). (4)
slow pion charge; this method tags the flavor of charged $D^*$ from
the decay of the $B$ and thus its flavor. The methods are applied
sequentially. The efficiency and wrong tag fractions are determined
from Monte Carlo studies and from analysis of a sample of self-tagging
exclusively reconstructed $B\rightarrow D^{(*)}l\nu$ decays. The
effective efficiencies ($\epsilon_{eff}=(1-2\omega)^2$) range from
10.5\% to 0.7\% and with a total of about 22\%. The number of
reconstructed $B$ mesons which were flavor-tagged are listed in Table
\ref{jrod:tbl:Bnum}.

\subsection{Vertexing and extraction of \sinphi $^3$}

\begin{figure}[t]
\vspace*{13pt}
\centerline{\epsfig{file=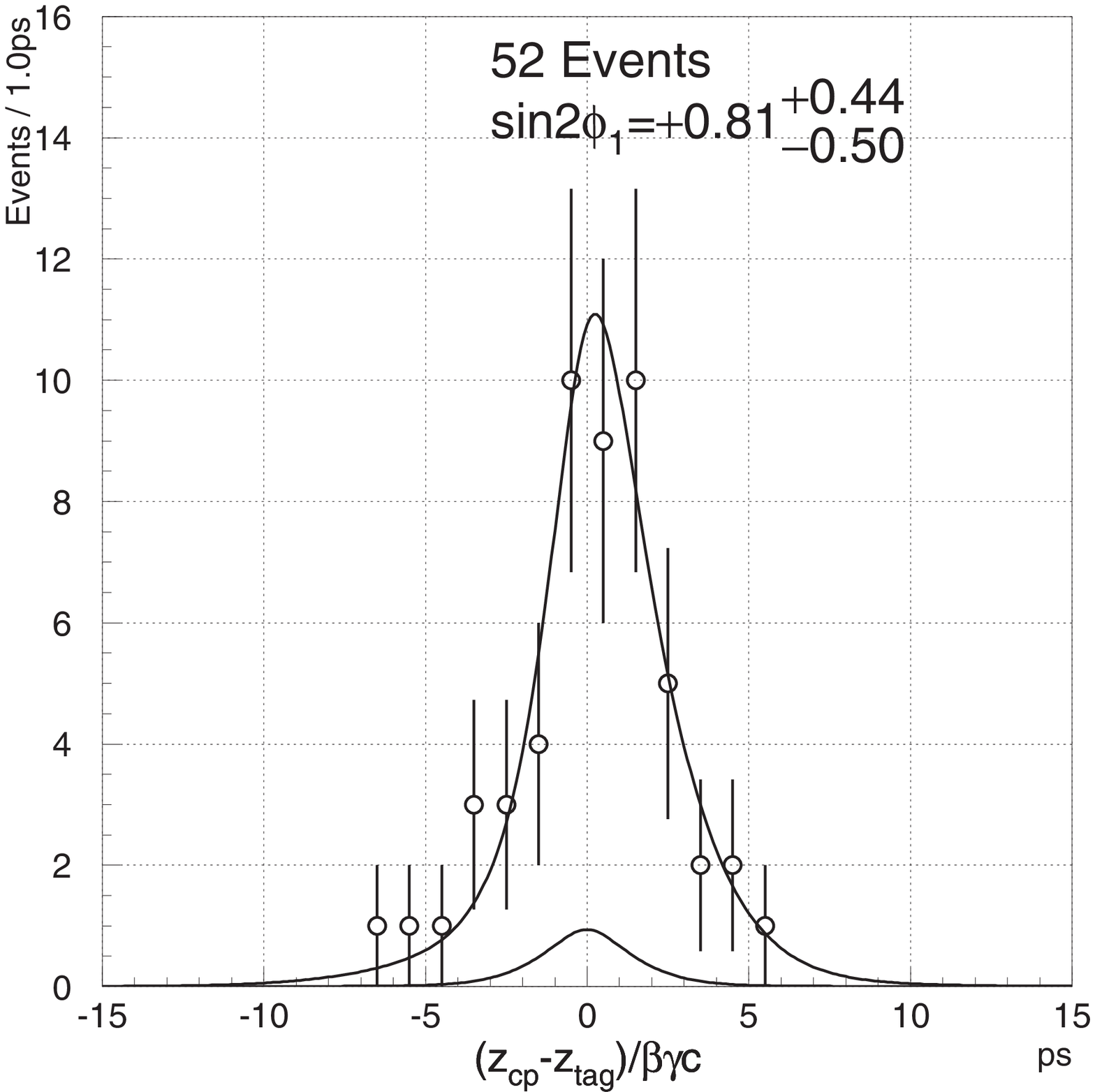,height=2.36in}\epsfig{file=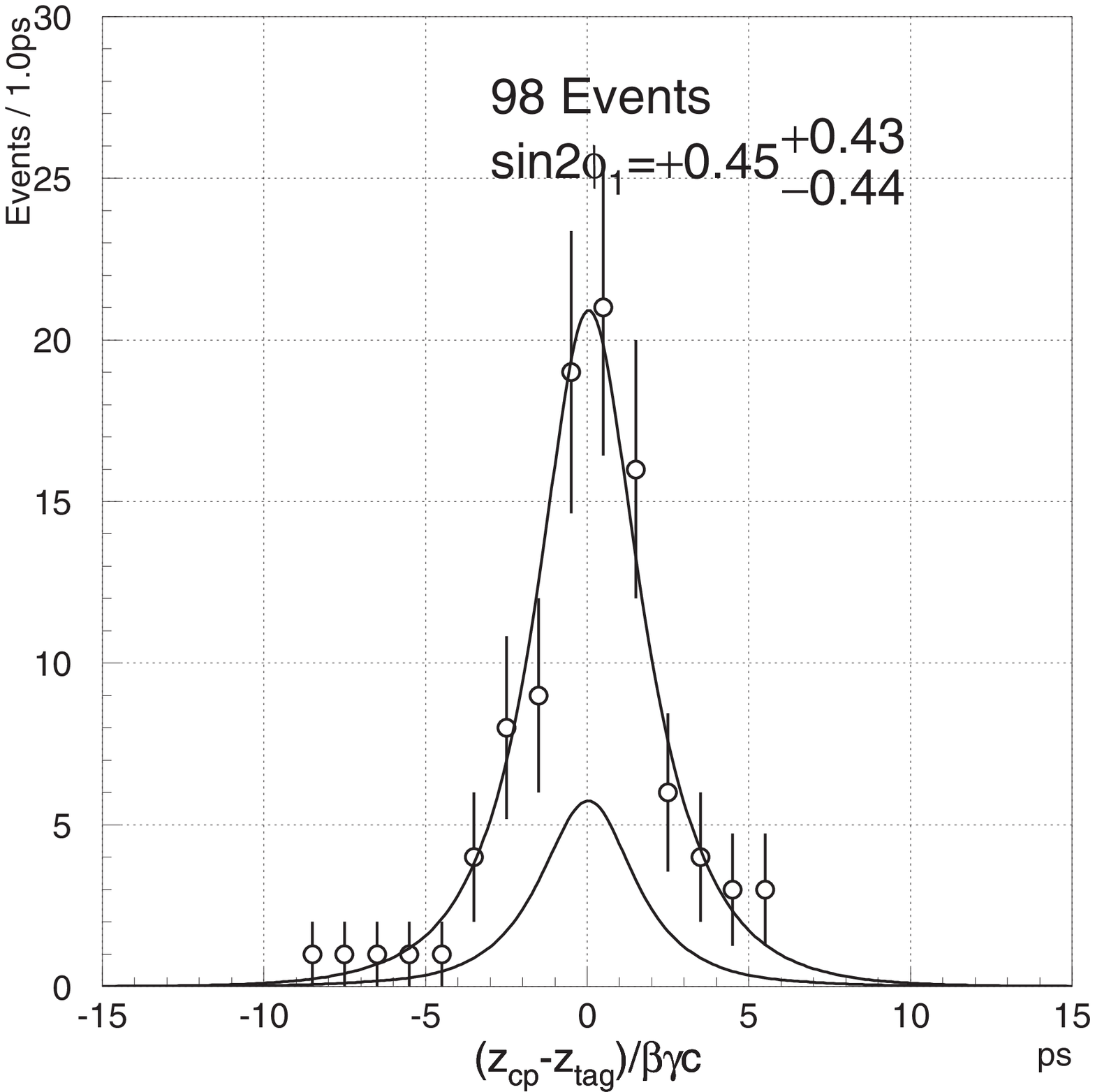,height=2.36in}}
\fcaption{The fitted proper-time distribution including data from both
$B^0$ and $\bar{B}^0$ candidates ($dN/d\Delta t|_{B^0} + dN/d(-\Delta
t)|_{\bar{B}^0}$). The plot on the left includes all modes with odd
$CP$. The plot on the right includes all modes.}
\label{jrod:fig:cpfits}
\end{figure}
\noindent
The value of \sinphi is extracted from an unbinned maximum-likelihood
fit to the proper-time distribution of tagged and fully reconstructed
$B_{CP}$ decays. The proper-time for each candidate event was
estimated from $\Delta z/\beta\gamma c$ where $\Delta z$ is the
difference between the decay vertices of the $B_{CP}$ and
$B_{tag}$. The vertex position of the $B_{CP}$ was established by the
two tracks assigned to the $J/\psi$ candidate. The vertex position of
the $B_{tag}$ was determined from tracks not assigned to $B_{CP}$ by
an algorithm that removes tracks which originate in secondary vertices
or do not contribute positively to the vertex fit.

To improve the statistical accuracy of our measurement the data from
all decay modes including those with odd and even $CP$ were combined
in the final likelihood fit. The likelihood function for each event
takes into account the finite resolution of the detector, the
charm-lifetime and includes terms for background contributions from
decays with or without their own $CP$ asymmetry. The $CP$-asymmetry,
$B^0\bar{B}^0$ mixing and dilution due to wrong-tagged events were
included in the functional form as illustrated by Equation~1.  The
maximum-likelihood fits to the data are shown in
Fig.~\ref{jrod:fig:cpfits}. Our preliminary result is:
\begin{equation}
\sin 2\phi_1 = 0.45^{+0.43}_{-0.44}({\rm stat})^{+0.07}_{-0.09}({\rm sys}).
\end{equation}

The systematic error includes effects from uncertainty in the fraction
of wrong tags, the $\Delta t$ resolution function for both background
and signal and input values used for $B$ lifetime and mixing
parameter. The dominant effect comes from the uncertainty in the wrong
tagged fraction. We also checked for biases in the analysis by
examining modes where no $CP$-asymmetry is expected. In particular we
ran the entire analysis including the vertexing, tagging and
$CP$-fitting algorithm on reconstructed samples of $B\rightarrow
J/\psi K^{*0}(K^-\pi^+)$,$B^-\rightarrow J/\psi K^-$, $B^-\rightarrow
D^0\pi^-$ and $B^0\rightarrow D^{*-}l^+\nu$. We observed no
$CP$-asymmetry in these modes as expected.

\section{Conclusion}
We have obtained a preliminary value for the \sinphi by analyzing
6.2 $fb^{-1}$ of data collected at the $\Upsilon(4S)$. Due to the
large statistical errors our result does not, at this time, show
conclusive evidence for $CP$ violation in the $B$ system. As more data
become available we expect to significantly improve the accuracy of
our measurement and establish whether or not the Standard Model can
account for $CP$ violation in the $B$ system.

\end{document}